\documentclass[useAMS,usenatbib,usegraphicx]{mn2e}

\usepackage{psfig}
\usepackage{times}
\usepackage{subfigure}

\newcommand{\msun}{\,\mbox{$M_{\odot}$}}

\newcommand{\kms}{\hbox{km s$^{-1}$}}
\newcommand{\ms}{\hbox{m s$^{-1}$}}

\newcommand{\vsini}{\hbox{$v$\,sin\,$i$}}

\newcommand{\vrad}{\hbox{$v_{rad}$}}
\newcommand{\degs}{$\degr$}
\newcommand{\chisq}{$\chi^{2}$}

\newcommand{\radday}{\hbox{rad.day$^{-1}$}}

\newcommand{\ha}{H$\alpha$}

\newcommand{\speedy}{\hbox{Speedy Mic}}

\begin{document}

\title[Speedy Mic]{The highly spotted photosphere of the young rapid rotator Speedy Mic}

\makeatletter
 
\def\newauthor{%
  \end{author@tabular}\par
  \begin{author@tabular}[t]{@{}l@{}}}
\makeatother

\author[J.R.~Barnes]
{J.R.~Barnes$^1$\thanks{E-mail: jrb3@st-andrews.ac.uk}\\
$^1$ School of Physics and Astronomy, University of St Andrews, Fife KY16 9SS. UK. }

\date{2005, 200?}

\maketitle

\begin{abstract}
We present high resolution images of the young rapidly rotating K3 dwarf \speedy\ (\hbox{BO Mic}, \hbox{HD 197890}). The photospheric spot maps reveal a heavily and uniformly spotted surface from equatorial to high latitude regions. Contrary to many images of similar objects, \speedy\ does not possess a uniform filling at high latitudes, but exhibits structure in the polar regions showing greatest concentration in a particular longitude range.

The asymmetric rotation profile of \speedy\ indicates the presence of a companion or nearby star which shows radial velocity shifts over a timescale of several years. Using a simple dynamical argument, we show that \speedy\ is unlikely to be a binary system, and conclude that the feature must be the result of a chance alignment with a background binary.

Complete phase coverage on two consecutive nights in addition to 60\% phase coverage after a three night gap has enabled us to track the evolution of spots with time. By incorporating a solar-like differential rotation model into the image reconstruction process, we find that the equator laps the polar regions once every \hbox{$191 \pm 17$ d}. This finding is in close agreement with measurements for other late-type rapid rotators.

\end{abstract}


\section{INTRODUCTION}

The distribution of starspots on young rapidly rotating stars has been determined for a number of objects \citep{strassmeier02survey}; and in some instances we now have photospheric image reconstructions at several epochs on a single object. While patterns in the distribution of emergent flux, as a function of spectral type for example, are just becoming discernible, it is still not entirely clear how the distribution may change on a single object with time. Direct evidence for chromospheric variation on timescales similar to the solar magnetic activity cycle have shown that some moderately rotating stars possess similar cycles of their own \citep{wilson78,baliunas95,donahue96}. Similarly, period variations derived from photometry \citep{hall91dynamo,henry95diffrot,messina02} have indicated that rapid rotators also show periodic variation in the emergent latitude of starspots. While long term variation in photometric brightness indicates cyclic activity, no evidence for variation of emergent latitudes in Doppler images has yet been presented, since very few objects have been observed consistently over a long enough time basis.

Here we present new observations of one of the fastest and brightest young rapidly rotating stars in the solar neighbourhood. \citet{montes01kinematic} have shown that Speedy Mic, like its rapidly rotating cousins AB Dor and PZ Tel is indeed a member of the Local Association. Speedy Mic (HD197890) is a K3 dwarf star with an equatorial rotation velocity, \vsini\ $\simeq$ 128 \kms. We first observed this star in 1998 \citep{barnes01speedy_b01} (B01), finding that spots are distributed at a range of latitudes. In B01 we gave a summary of various observations of Speedy Mic. Since the first detection of magnetic activity \citep{bromage92speedy,matthews94speedy}, it has been shown that Speedy Mic is one of the most active solar neighbourhood stars \citep{singh99xray}, with an X-ray luminosity of {  Log}(\mbox{$L_x/L_{bol}$}) = -3.07. \citet{singh99xray} also found evidence for flaring activity and significantly subsolar coronal Fe abundances in accordance with observations of other active stars. A more recent measurement by \citep{marakov03xray} indicated an X-ray luminosity of \hbox{ \mbox{$L_x$} = $116.78 \times 10^{29}$ ergs s$^{-1}$}, an order of magnitude higher than that of the well studied K0 star AB Dor \hbox{ ($L_x = 15.22 \times 10^{29}$ ergs s$^{-1}$)} for example. This implies { Log}(\mbox{$L_x/L_{bol}$}) $\sim$ -2, which is an order of magnitude higher than measurements for similar stars, again indicating probable flaring activity at the time of observations.

\begin{table*}
\begin{minipage}{135mm}
\caption{{ Observation log for 2002 July 18, 19 \& 23}}
\protect\label{tab:obs_journal}
\vspace{5mm}
\begin{center}
\begin{tabular}{lccccl}
\hline
Object          & UT Start& UT End& Exp time    & No. of frames	& comments \\
		&	&	&  [s]		&		&	 \\
\hline
		& 	&	& 2001 Jul 18	&              	&	 \\
\hline
HD106911	& 08:41	& 08:42	&	40	& 1		& B5V telluric standard \\
HD126053	& 08:46	& 08:51	&	250	& 1		& G1V RV standard 	\\
Gl 673		& 08:53	& 08:58	&	300	& 1		& K5V spectral standard	\\
Gl 472		& 09:01	& 09:06	&	300	&		& K1V spectral standard	\\ 
Gl 701		& 09:57	& 10:07	&	600	&		& M1V spectral standard	\\ 
Speedy Mic	& 10:10	& 19:42	&	300	& 102		& Target \\
Gl 849		& 19:45	& 19:58	&	800	&		& M3.5V spectral standard \\ 
HD 16160	& 20:01	& 10:06	&	300	&		& K3V spectral standard	\\ 
\hline
		& 	&	& 2002 Jul 19	&              	&			\\
\hline
HD106911	& 08:07	& 08:08	&	60	& 1		& B5V telluric standard \\
HD119850	& 08:14	& 08:24	&	600	& 1		& M1.5V spectral standard \\
Speedy Mic	& 10:05	& 19:18	&	300	& 97		& Target 		\\
Gl 825		& 19:19	& 19:24	&	300	& 1		& K7V spectral standard	\\ 
HD 4628		& 19:26	& 19:29	&	180	& 1		& K2V spectral standard	\\ 
\hline
		& 	&	& 2002 Jul 23	&              	&			\\
\hline
HD106911	& 08:02	& 08:03	&	40	& 1		& B5V telluric standard \\
Speedy Mic	& 09:46	& 18:54	&	300	& 97		& Target 		\\
Gl 825		& 19:19	& 19:24	&	80	& 1		& K7V spectral standard	\\ 
HD 4628		& 19:26	& 19:29	&	180	& 1		& K2V spectral standard	\\ 
\hline

\end{tabular}
 \end{center}
\end{minipage}
\end{table*}

In B01 we also confirmed the presence of \ha\ transients, first reported by \citet{jeffries93speedyprom} . These transients are thought to be the result of clouds of cool material, analogous to solar prominences, passing at large distances from the rotation axis in front of the stellar disc. Both studies found prominences at or above the co-rotation radius of the star. The \ha\ profile in the present data set is studied in a separate publication \citep{dunstone05speedy}, revealing a densely packed prominence system beyond the co-rotation radius.

In this paper we present images at closely separated epochs. We examine the reliability of the strength of reconstructed features in the maps through use of a bootstrapping method. This allows us to distinguish between real evolution of starspot features and uncertainties in the reconstruction process. We compare our images with those published by \cite{wsw05speedy} (WSW05), derived from observations made between two and three weeks after those presented here. We also combine the complete timeseries and determine the differential rotation rate at the photospheric level.

\section{OBSERVATIONS}

Observations were made with the University College London \'{E}chelle Spectrograph (UCLES) at the Anglo Australian Telescope (AAT) on 2002 July 18, 19 \& 23. With a \mbox{1.2\arcsec}~slit width, a mean spectral resolution of $\sim$ 44000 at 5450\AA~was attained with coverage from 4376 \AA\ to 6892 \AA\ in 47 orders on the 2Kx4K EEV CCD. This corresponds to $\sim$ 6.7 \kms\ at the { mean} wavelength. { The pixels were binned by a factor of two in the wavelength direction giving two pixels per spectral resolution element. Exposure times for Speedy Mic were 300 secs for all spectra.} The seeing varied between 0.9\arcsec\ and 2\arcsec\ with a mean of $~$1.5\arcsec. Cloud gradually built up to 80\% cover during the second half of July 23. This resulted in loss of target counts and contamination of spectra by moonlight. We have corrected for this in a similar manner to that described in \citet{donati00rxj1508} and \citet{barnes01mdwarfs}. { We do not detect any further significant changes in continuum level during the course of observations, which may be due to flaring activity for instance.}

\begin{figure*}
\begin{center}

  \begin{tabular}{ccc}
    \includegraphics[width=5.0cm,height=10cm,angle=0]{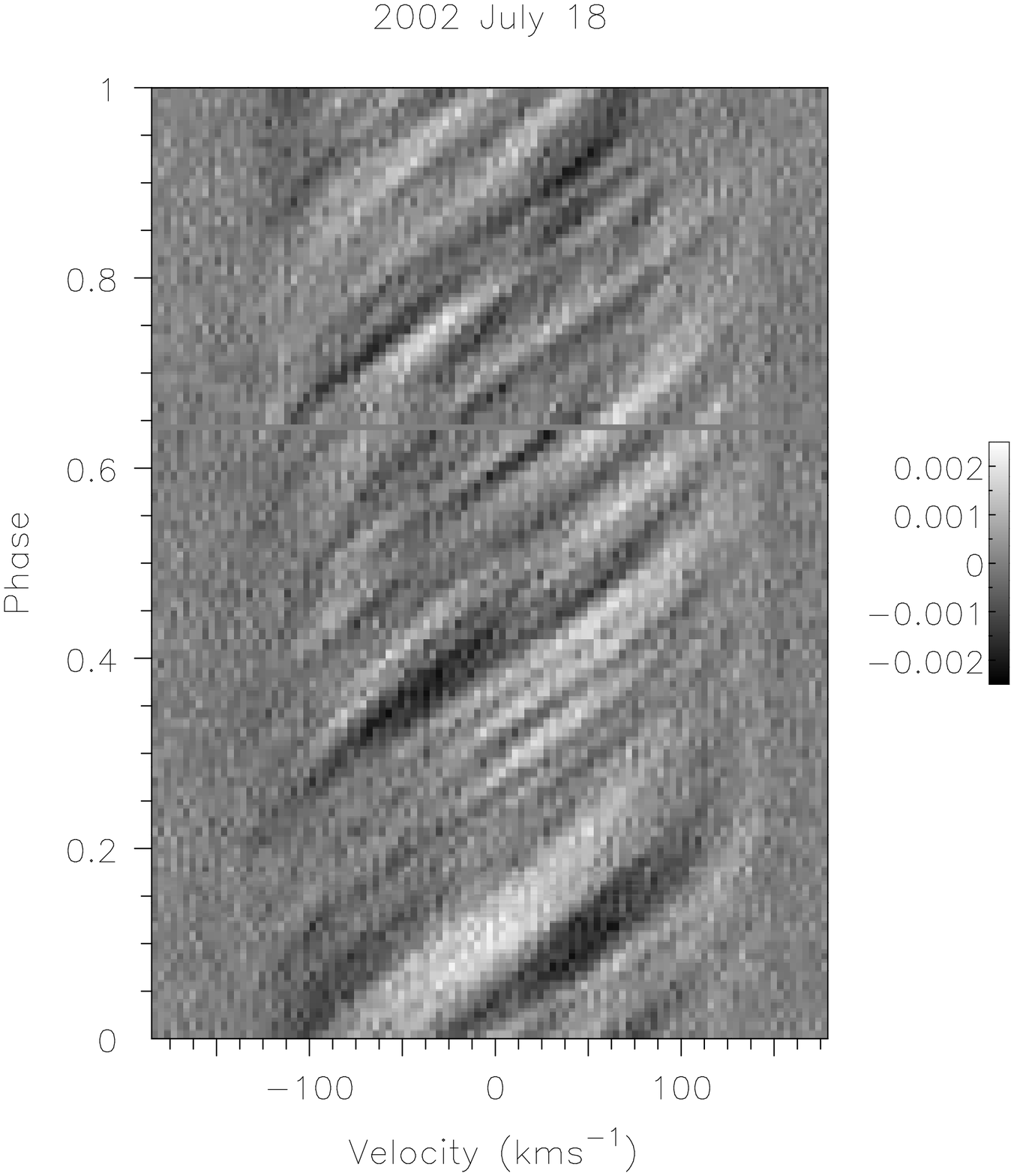} &
    \hspace{-2mm}
    \includegraphics[width=5.0cm,height=10cm,angle=0]{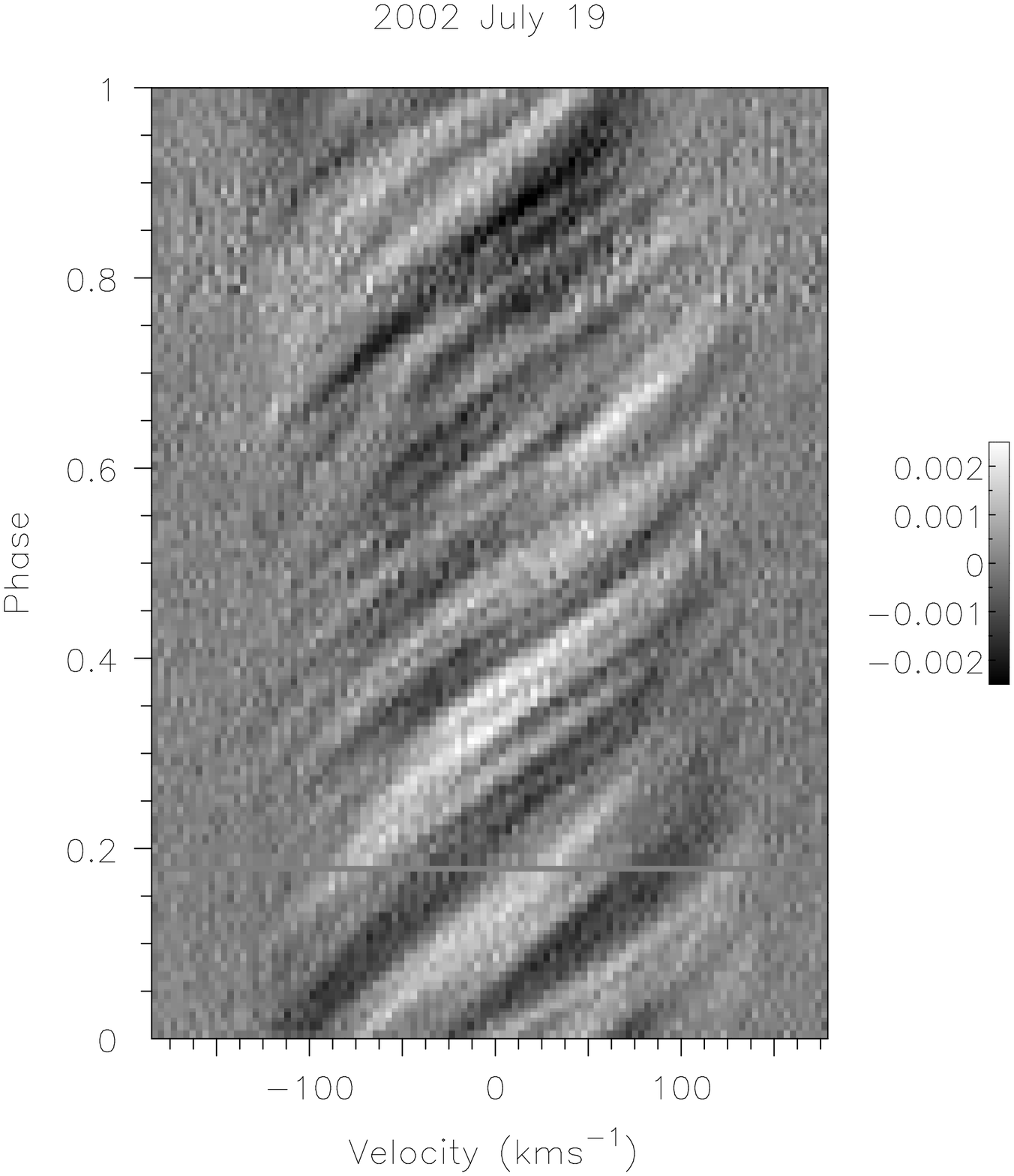} &
     \hspace{-2mm}
    \includegraphics[width=5.0cm,height=10cm,angle=0]{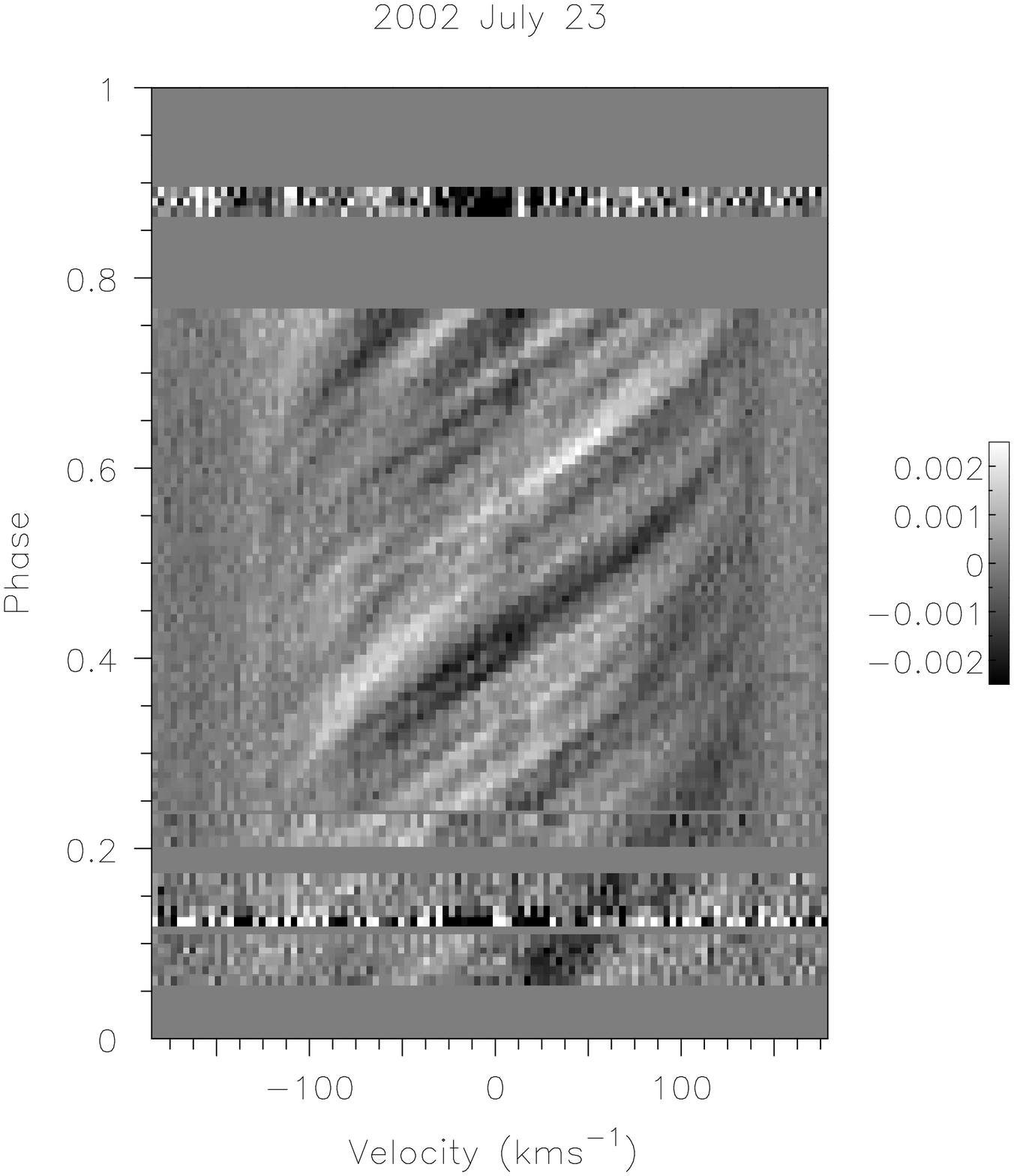} \\
  \end{tabular}

\end{center}
\caption[Speedy Mic time series spectra ]{Time series spectra for 2002 July 18, 19 \& 23, with phase plotted against velocity. The mean profile has been subtracted from each time series. White features correspond to starspot signatures. }
\protect\label{fig:timeseries}
\end{figure*}

We observed a number of slowly rotating template stars with effective temperatures similar to the photosphere and spots of Speedy Mic. These spectral standards are used to represent the local intensity profiles in the image reconstruction process. A featureless rapidly rotating B5V star (HD 106911) was observed as a telluric standard and also served the purpose of tracing the spectral orders in the extraction process.

\section{DATA REDUCTION}

\subsection{Extraction}

Pixel to pixel variations were removed using flat-field exposures taken with an internal tungsten reference lamp. The worst cosmic ray events were removed at the pre-extraction stage using the FIGARO routine {\sc bclean} \citep{shortridge93figaro}. Scattered light was modelled by fitting polynomials of degree 7 to the sets of inter-order pixels at each X-position in each frame. The spectra were extracted using ECHOMOP, the \'{e}chelle reduction package developed by \citet{mills92}. The Thorium-Argon arc-frames used for wavelength calibration were extracted in conjunction with a target spectrum, and calibrated using this package. The orders were extracted using ECHOMOP's implementation of the { optimal} extraction algorithm developed by \citet{horne86extopt}. ECHOMOP propagates error information based on photon statistics and readout noise throughout the extraction process. \\

\begin{table}
\caption{Mean S/N statistics for each night of observations. Cloud was responsible for poorer S/N ratio in the latter part of the night of 2002 July 23. We thus give statistics in two parts for this night.}
\protect\label{tab:stats}
\begin{center}
\begin{tabular}{ccccc}
\hline
Date	& No. of spectra&     	Input S/N ratio	& Output S/N ratio	& Gain     \\
\hline
18	&	100	&	77.0 $\pm$ 5.5	&	3729 $\pm$ 493	&	48 \\
19	&	97	&	74.3 $\pm$ 11.5	&	3692 $\pm$ 766	&	50 \\
23 	&	51	&	86.8 $\pm$ 4.4	&	4211 $\pm$ 438	&	49 \\
23	&	16	&	43.7 $\pm$ 14.8	&	1784 $\pm$ 969	&	41 \\
\hline
\end{tabular}
\end{center}
\end{table}

\section{THE SPECTRA}

We applied least squares deconvolution \citep{donati97zdi, barnes98aper} to each spectrum, thereby deriving a single profile at each observation phase. The list of lines used for deconvolution was taken from the Vienna Atomic Line Database \citep{kupka99} for a star with T = 4750 K and log g = 4.5. In total, 5680 lines with depths between 0.05 and 1.0 were used in the deconvolution. Due to overlap of spectral coverage of adjacent orders in the blue part of the spectrum, a total of 7547 images were used to derive the rotation profile. The noise statistics are presented in Table \ref{tab:stats}. The phased timeseries spectra are plotted in Fig. \ref{fig:timeseries} where the mean profile has been subtracted to reveal many starspot (white trails) signatures.

\subsection{System parameters}
\protect\label{parameters}

The system parameters for Speedy Mic have been determined by a number of authors, including B01, in a previous paper based upon spectra taken during 1998 July. During the 1998 observations, only a few spectra at a number of different instrumental configurations (at two telescopes) were available. However the data set presented here provides us with nearly three complete rotation cycles with very high quality data. Using our Doppler imaging code, we combined all 264 usable spectra into a single timeseries and re-determined the radial velocity, $v_{rad}$, inclination, $i$, equatorial rotation velocity, \vsini, and period, $P$. The period measurement does not include differential rotation shear which is measured and discussed in a later section. We also emphasise that the period as measured relies on the repetition of the starspot trails in the timeseries, from night to night, and thus represents the mean period of Speedy Mic. The quoted uncertainty is for the measurement on this data set and does not take into account any changes which may occur if the mean rotation rate changes. In other words, if a magnetic cycle is present then the mean location of spots may change leading to a different rotation period measurement. The period is nevertheless in agreement with the 0.380 $\pm$ 0.004 d photometric period found by \cite{cutispoto97speedy} 

\begin{figure}
\begin{center}
\includegraphics[height=8cm,angle=270]{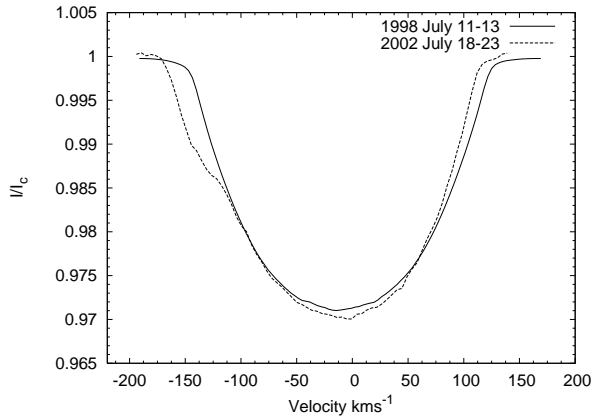}
\end{center}
\caption{Mean profile from all spectra. The 1998 July 11-13 profile is plotted with the mean profile derived from observations in 2002 July 18, 19 \& 23.}
\protect\label{fig:meanprofile}
\end{figure}

As with the 1998 data set, we deconvolved the spectrum of the slowly rotating K3 star HD 16160 in the same manner as our target spectra. This profile was used to represent the local intensity profile in the imaging process, and was scaled according to foreshortening cosine and limb darkening. We used a 4-parameter limb darkening law \citep{claret00ldc4} interpolated to a temperature of T = 4890 K. 

\begin{table}
\caption[System parameters]{System parameters for Speedy Mic}
\protect\label{tab:sysparam}
\vspace{5mm}
\centering
\begin{tabular}{lc}
\hline
P [d]			&  0.38007 $\pm$ 0.00005   \\
$v_{r}$~[kms$^{-1}$]	&  -8.0 $\pm$ 1.0   \\
\vsini 	~[kms$^{-1}$]	&  132 $\pm$ 2 \\
Axial inclination [deg]	&  70.0 $\pm$ 5   \\
\hline
\end{tabular}
\end{table}

The mean profile from all spectra indicates a significant asymmetry in the rotation profile in that the red wing of the profile exhibits excess absorption. We discuss this feature at length in \S \ref{sec:binary} in light of this finding and previous high resolution spectroscopic observations made on Speedy Mic. Since we are certain that this absorption feature does not arise from Speedy Mic itself, we have excluded it from the imaging process by artificially inflating the error bars (effectively to infinite size) in the region of \hbox{85 \kms}\ to \hbox{150 \kms}. The  system parameters are presented in Table \ref{tab:sysparam} and indicate values which are close to our original estimations. Notably, the preferred inclination is $i$ = 70\degs\ rather than the $i$ = 55\degs\ which we initially measured. The \vsini\ measurement is 3\% greater than that measured on the 1998 data set. We attribute this to removal of the redshifted absorption feature which effectively removes the right hand portion of the profile. This includes the region of strong curvarture where the line meets the continuum, normally providing a strong constraint on the extent and therefore \vsini\ of Speedy Mic. Since we had to perform a similar task on the blue wing of the 1998 data set, where the absorption feature appeared even more significant, it is likely that there is some subjective element in choosing which region to exclude from the fit. If too much is excluded, we are likely to slightly underestimate the \vsini\, while excluding too little may overestimate \vsini. Given the higher and more consistent quality of the current data set we will assume \hbox{\vsini\ = 132~\kms}\ in this paper. The 3\% difference in \vsini\ will nevertheless have little effect on the image reconstructions.

\begin{figure*}
\subfigure[]{\includegraphics[height=152mm,angle=270,bbllx=70,bblly=0,bburx=355,bbury=800]{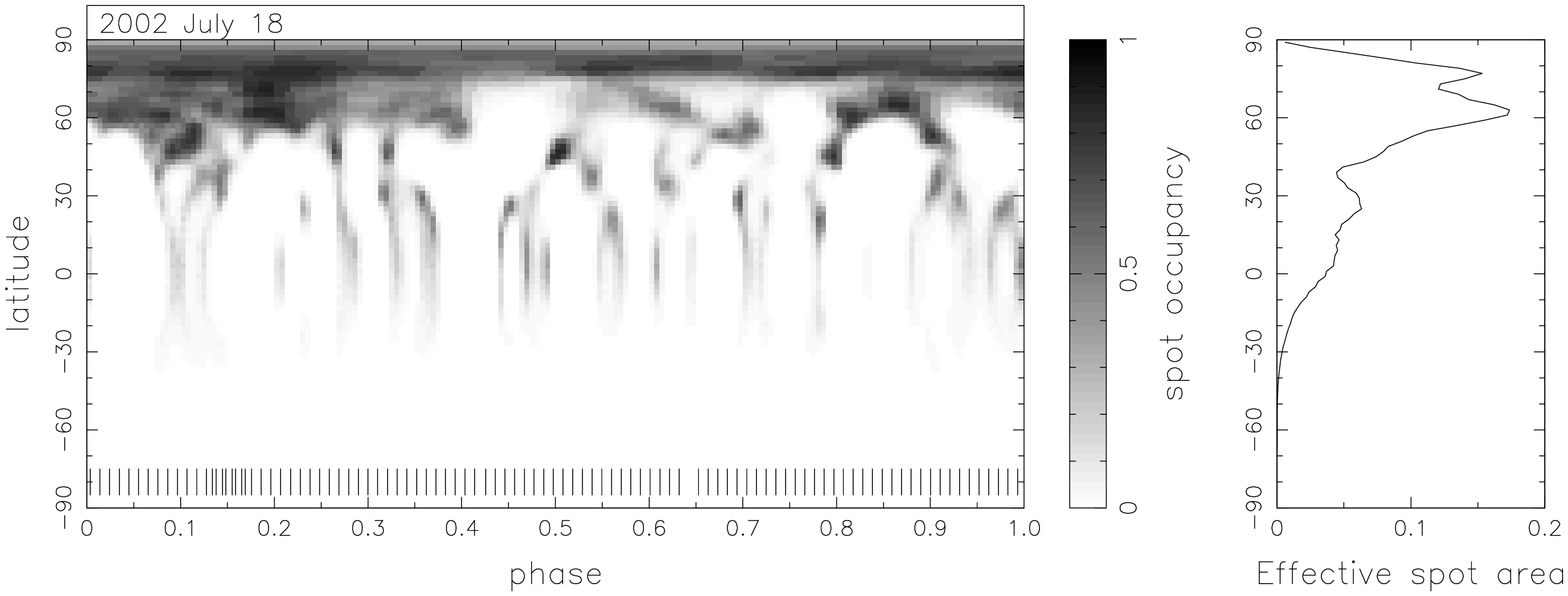} }
\subfigure[]{\includegraphics[height=152mm,angle=270,bbllx=70,bblly=0,bburx=355,bbury=800]{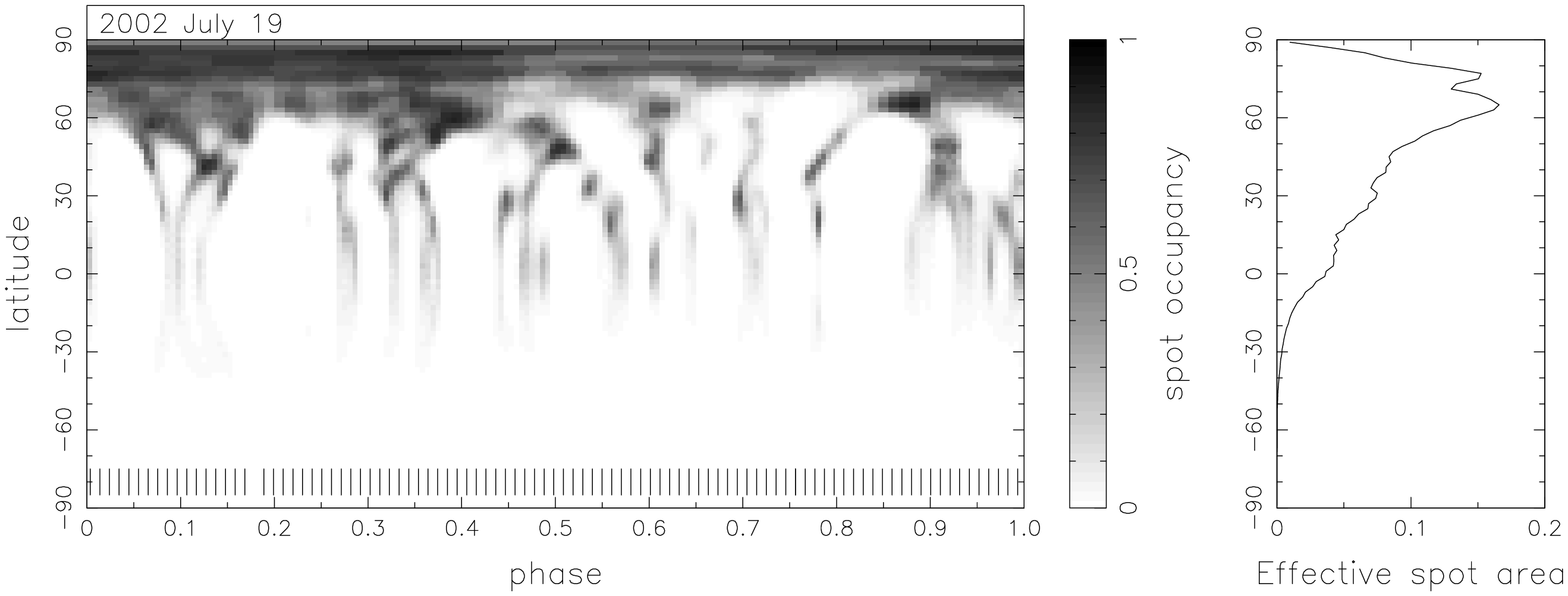} }
\subfigure[]{\includegraphics[height=152mm,angle=270,bbllx=70,bblly=0,bburx=355,bbury=800]{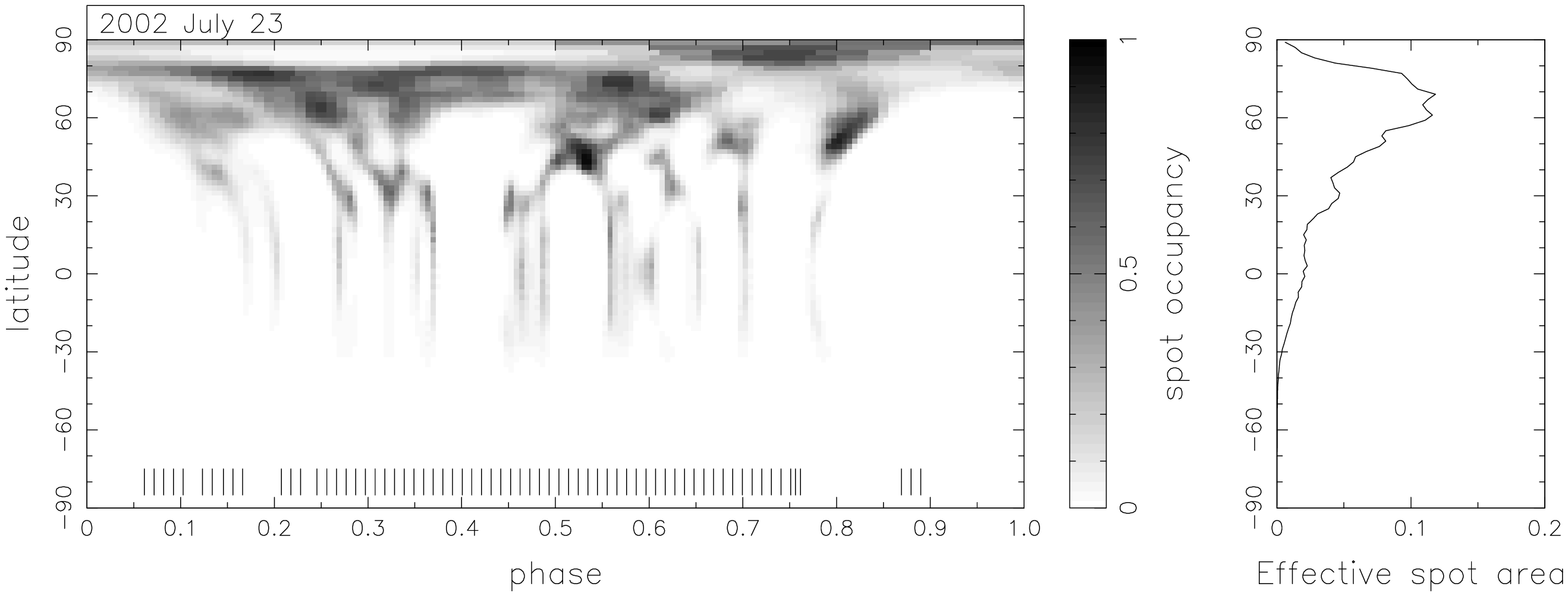} }
\caption[Speedy reconstructed images]{Maximum entropy regularised reconstruction of \speedy\ for 2002 July (a) 18, (b) 19 and (c) 23.}
\protect\label{fig:speedy_images}
\end{figure*}

\begin{figure*}
\subfigure[]{\includegraphics[height=152mm,angle=270,bbllx=70,bblly=0,bburx=355,bbury=800]{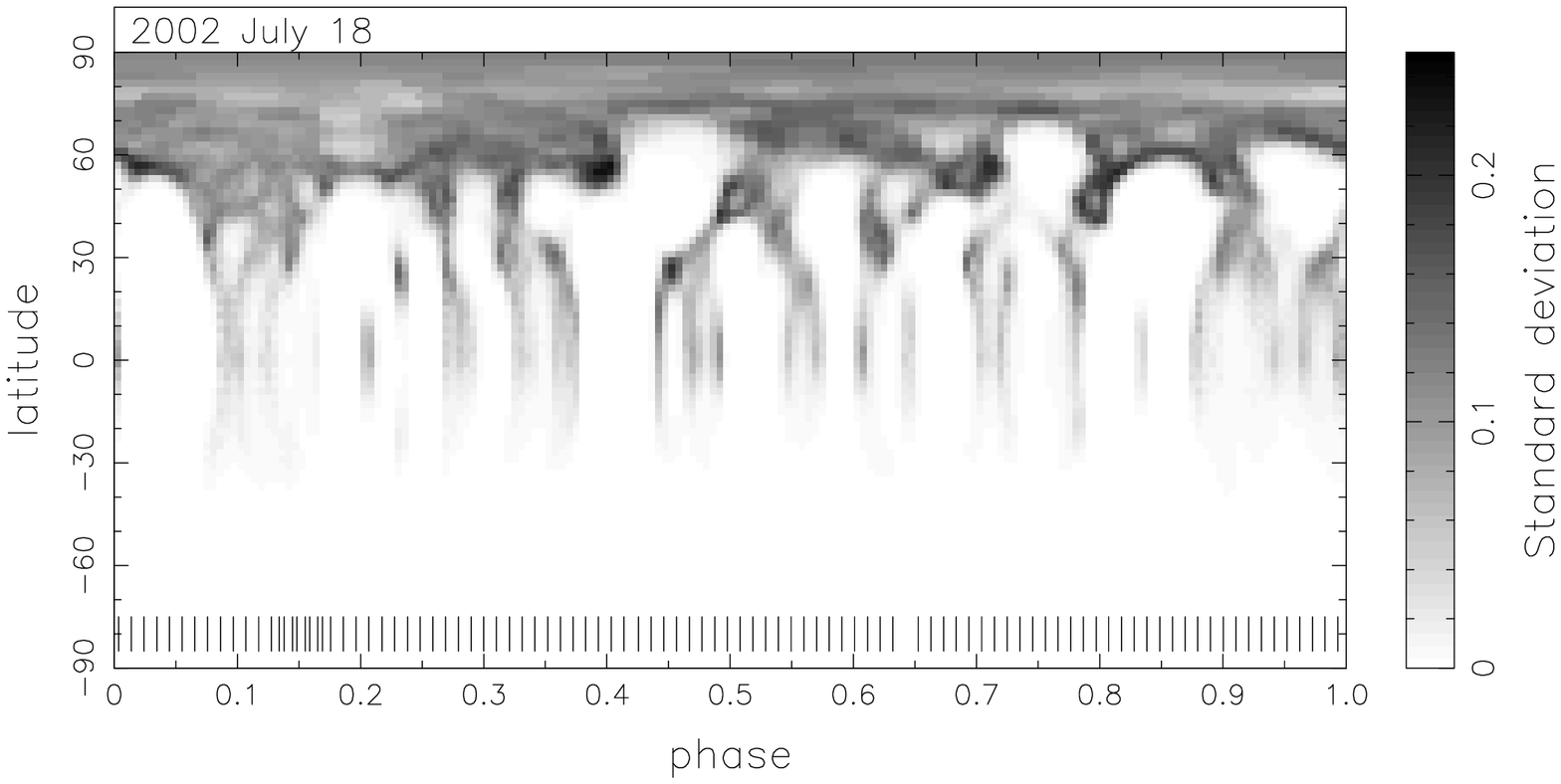} }
\subfigure[]{\includegraphics[height=152mm,angle=270,bbllx=70,bblly=0,bburx=355,bbury=800]{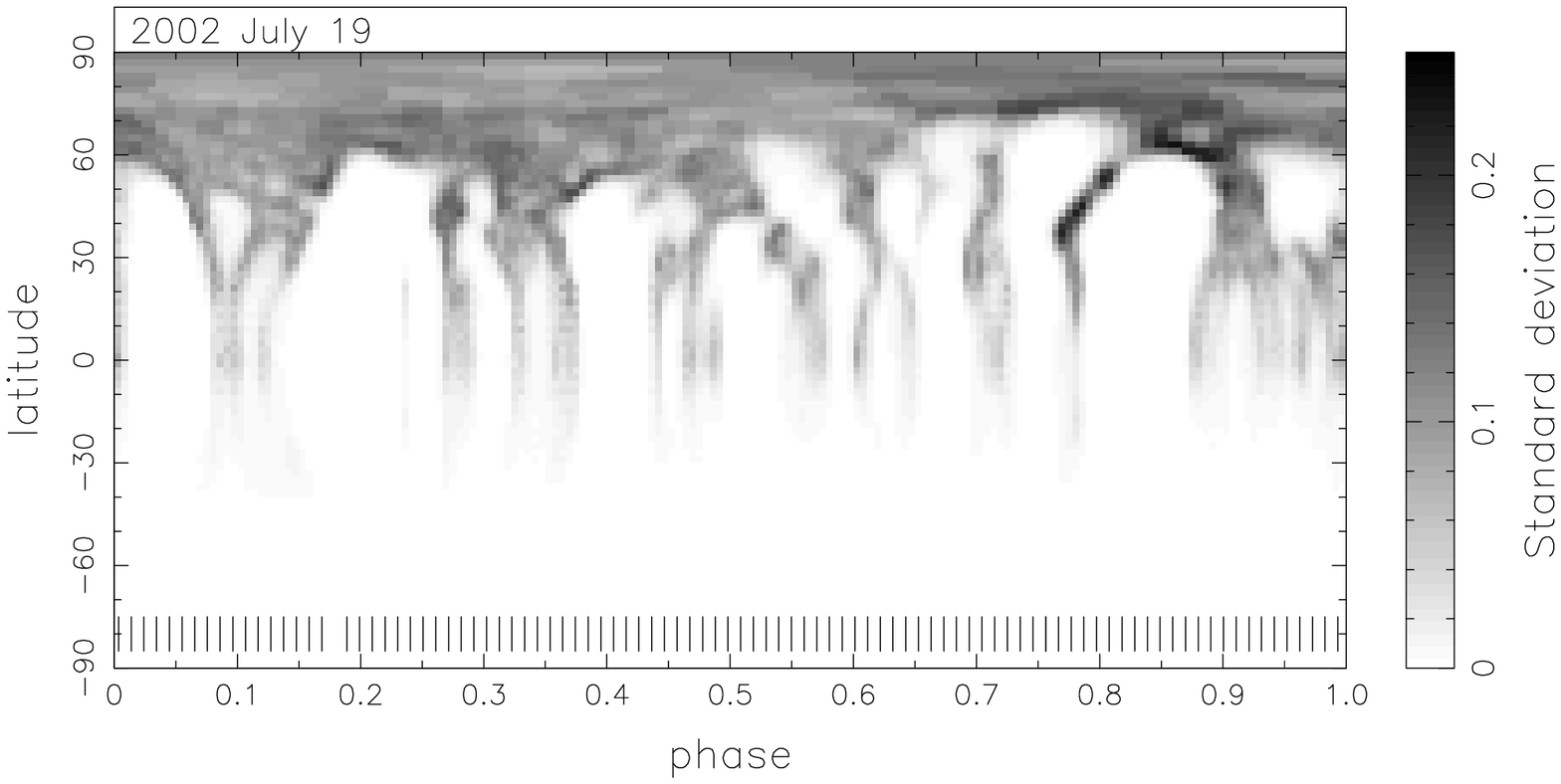} }
\subfigure[]{\includegraphics[height=152mm,angle=270,bbllx=70,bblly=0,bburx=355,bbury=800]{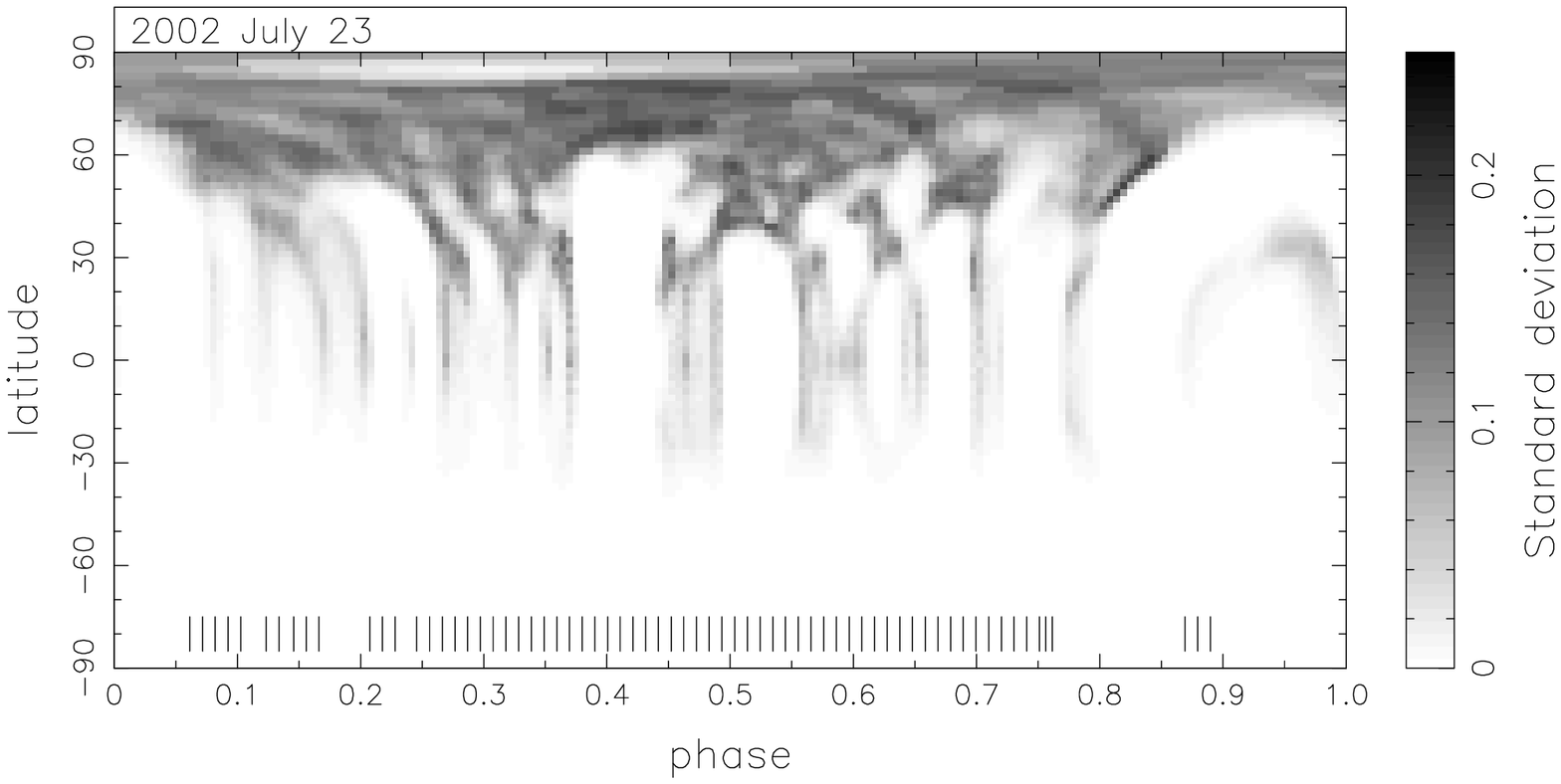} }
\caption[Speedy Variance Maps]{Standard deviation maps from bootstrap reconstructions of 100 data sets. Note that the range of plotted values is 0 to 0.25 (plotted for clarity), so the greyscale is not directly comparable with that used in Fig. \protect\ref{fig:speedy_images}.}
\protect\label{fig:speedy_stddev}
\end{figure*}

\section{SPEEDY MIC: SINGLE STAR OR BINARY?}
\protect\label{sec:binary}

\subsection{Real or artifact?}
\protect\label{sec:realartifact}

The mean profile in our timeseries indicates excess absorption in the red wing. A similar feature was present in the 1998 data (B01) which we attributed to the presence of a background star in the slit at the time of observations. The re-appearance of this feature in the red wing of the rotation profile (Fig. \ref{fig:meanprofile}) suggests the possible presence of a companion star, indicating that Speedy Mic may in fact be a binary or multiple system. In order to ascertain whether the absorption feature is real and not an artifact, or a problem with the deconvolution process, we deconvolved various combinations of blue and red orders of each spectrum independently for the night of \hbox{July 18}. In all cases, the excess absorption in the red wing is present. We also deconvolved the full spectrum using a slowly rotating template star rather than a line list in case mis-matches in the synthetic line list were causing problems. This method, described in \citet{barnes04svd}, uses singular value decomposition to perform the deconvolution. Here, it is possible to obtain a much closer match between the reconstructed spectrum (i.e. template re-convolved with rotation profile) and the observed spectrum. Applying this method of deconvolution nevertheless also resulted in the appearance of the absorption feature.

\subsection{Nearby objects}
\protect\label{sec:nearbyobjects}

Having confirmed that the absorption feature is real, we must consider that a nearby star was present in the slit of the spectrograph during the observations, or that Speedy Mic is in fact a binary or multiple system. We have searched the 2 { Micron} All Sky Survey (2MASS) \footnote{See http://irsa.ipac.caltech.edu} catalogue which does not list any objects close enough to Speedy Mic which could contaminate our spectrum. The 2MASS J-band image shows an object (unidentified by the survey) at a distance of approximately 11\arcsec\ and position angle of 13\degs. Compared with objects in the same field, it appears to possess a magnitude of J $\sim$ 13\,-\,14 and thus it seems unlikely that even if it is a binary system, it would contaminate our spectra. Similarly, the seemingly redder object which appears only in the K-band (K = 13\,-\,14) image at 14\arcsec\ and position angle 185\degs\ is unlikely to have appeared in the slit.

A survey for brown dwarf companions to nearby stars has been made by \citet{mccarthy04} using coronagraphs to occult the primary star. Below 5\arcsec\, the limit of detectability decreased almost linearly with distance, reaching a sensitivity of \hbox{$\sim$ $\Delta$J = 2} at 0.15\arcsec\ separation. Beyond 5\arcsec, the limiting \hbox{$\Delta$J = 12.5} remained approximately constant, however no candidate brown dwarfs were detected close to Speedy Mic in the survey.

\subsection{Is Speedy Mic a binary?}
\protect\label{sec:speedyabinary}

Observations of Speedy Mic in the past have attributed it single star status, nevertheless a range of rotation velocities have been reported. From the FWHM of the Ca {\sc ii} K line, \citet{bromage92speedy} estimated \hbox{\vsini\ = 120 $\pm$ 40 \kms}\ and \hbox{$v_{rad}$ = -6.5 \kms}\ with no systematic variations in the latter value over the six days during which observations were made. \citet{anders93speedy} however reported \hbox{\vsini\ = 170 \kms}\ using spectral synthesis of two lines centred around 6710~\AA. By combining data from simultaneous data at two observatories, B01 found that \hbox{\vsini\ = 128 \kms}\ and \hbox{$v_{rad}$ = -8.0 \kms.} A range of $v_{rad}$ of \hbox{-8.0 to -10.5} for individual nights of observations and instrumental setups was most likely due to incomplete phase coverage and inaccurate internal calibrations. The estimated error on this latter measurement is of order \hbox{2 \kms}. As was shown in \S \ref{parameters}, we found \hbox{\vsini\ = 132 \kms}\ and \hbox{\vrad\ = -8.0 \kms.}

We are therefore presented with the contradictory observations of no (or at most small) radial velocity variations and an absorption feature, apparently due to a secondary component, which appears in the blue wing at one epoch, and the red wing at another. To assess whether Speedy Mic could plausibly have a low mass companion, we have estimated the expected radial velocity amplitude of each component for a range of secondary stellar masses. With an assumed binary orbital inclination of 70\degs and a circular obit, the mass function is defined \citep{hilditchbook} as

\begin{equation}
f(m) = \frac{m_2^3 sin^3(i)}{(m_1 + m_2)^2} = 1.0361 \times 10^{-7}  K_1^3  P.
\end{equation}

where masses are given in \msun, period $P$ in days and $K_1$ in \kms. Table \ref{tab:masses} gives $K_1$ velocities for a range of secondary masses and periods. Since the absorption feature is not seen to move during the 5 night span of observations, we have assumed that the proposed star is at quadrature, and estimate that it cannot have moved through more than 10\% of the orbit. This yields a lower limit to the orbital period of 50 days. The maximum period of 8 years is assumed from the approximation that the absorption feature took 4 years (1998 to 2002) to traverse from the blue wing (B01) to the red wing (present observations).

\begin{table}
\caption{$K_1$ and $K_2$ velocities for a range of secondary masses and orbital periods of 50 d and $\sim$8 yrs. The primary mass is assumed to be 0.8 \msun. Secondary masses and spectral types estimated from \protect\citep{cox00allens}.}
\protect\label{tab:masses}
\begin{center}
\begin{tabular}{cccccc}
\hline
\multicolumn{2}{c}{Secondary}	& \multicolumn{2}{c}{$K_1$} 	 & \multicolumn{2}{c}{$K_2$} \\

Sp Type	& Mass [\msun]	 			& \multicolumn{4}{c}{P [days]}	 			 									\\
	&	$m_2$	& 50    	&     3000	&	50    	&     3000	\\
\hline	
M0	& 0.51		&	21.1	&	5.91	&	36.2	&	9.3	\\
M2	& 0.40		&	19.2	&	4.91	&	38.5	&	9.8	\\
M5	& 0.21		&	11.3	&	2.89	&	43.1	&	11.0	\\
M8	& 0.06		&	3.6	&	0.92	&	48.0	&	12.2	\\
\hline
\end{tabular}
\end{center}
\end{table}

Based on measurements made by various authors and our own measurements in (B01), we estimate an upper limit of 4 \kms\ to the radial velocity variations exhibited by Speedy Mic. This constrains an upper spectral type for the secondary component of M5 in the considered period range. In order that the secondary component exhibits sufficient radial velocity variation to explain the observed absorption features, we can narrow down the secondary spectral type to late M and a binary orbit of tens of days (the M8, $P$=50 d case in Table \ref{tab:masses}). The proposed secondary must itself be a rapid rotator and not tidally locked with the primary. However, the binarity argument fails because an M8/K5 pair would exhibit a large luminosity ratio, with $log(L_{\rm Speedy}/L_{\rm M8}) \sim -3.3$ \citep{golimowski04}, ensuring that the secondary profile would not be seen. 

The only remaining possibility is that \speedy\ is in a chance alignment with a binary. We find no evidence for starspot transients in this part of the profile, although they may be below the detection limit. We therefore conclude that the additional absorption feature arises from a background object which has remained undetected in present surveys due to its close proximity to \speedy.

\begin{figure*}
\begin{tabular}{c}
\includegraphics[height=152mm,angle=270,bbllx=70,bblly=0,bburx=355,bbury=800]{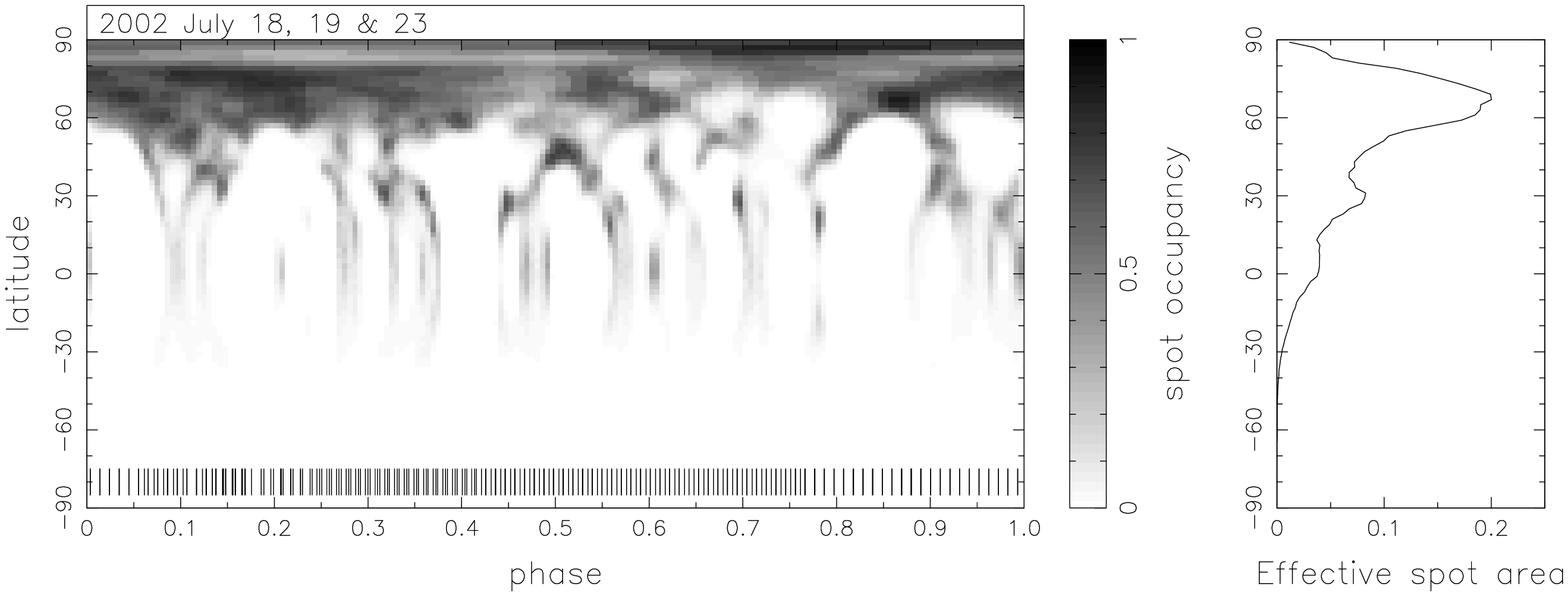} \\
\end{tabular}
\bigskip
\begin{tabular}{cccc}
\includegraphics[height=40mm,angle=0,bbllx=100,bblly=267,bburx=390,bbury=559]{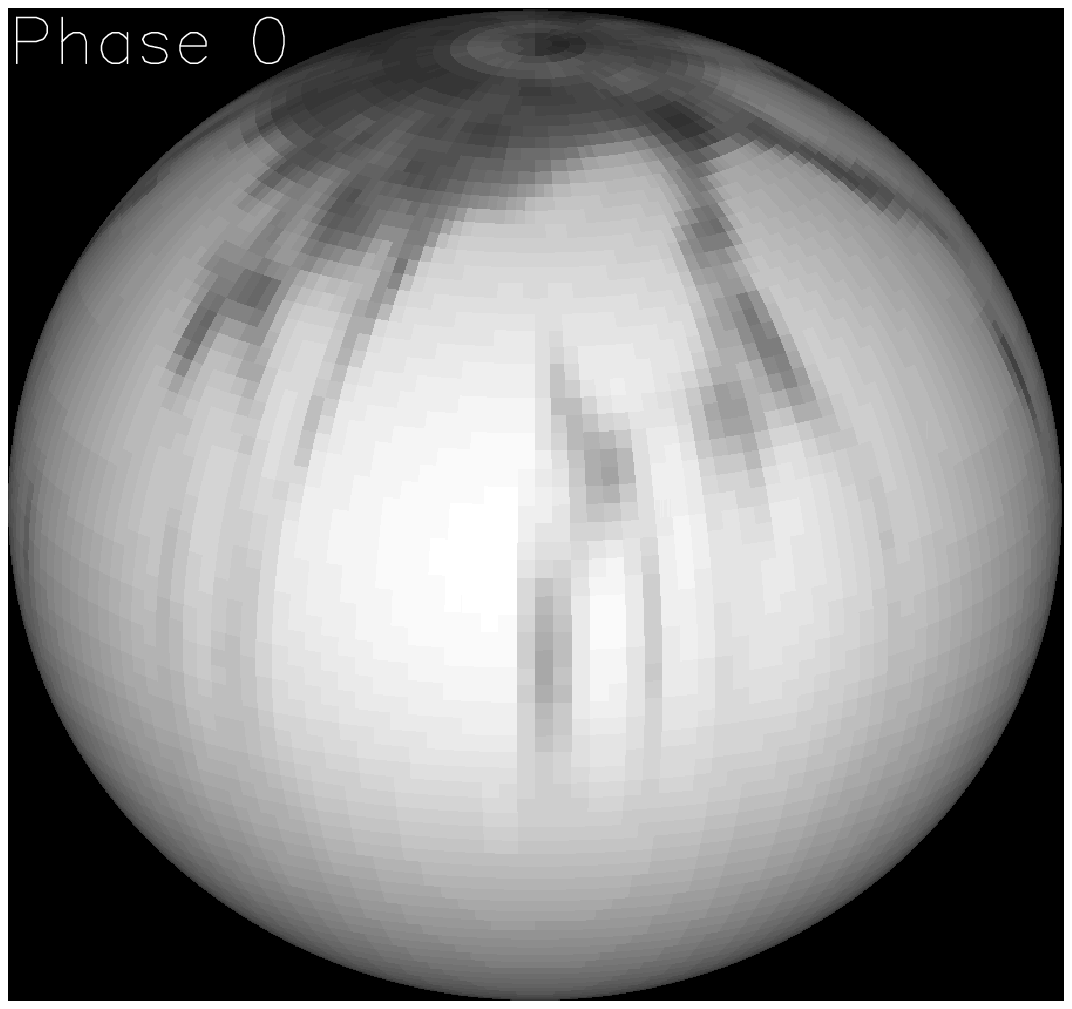} &
\includegraphics[height=40mm,angle=0,bbllx=100,bblly=267,bburx=390,bbury=559]{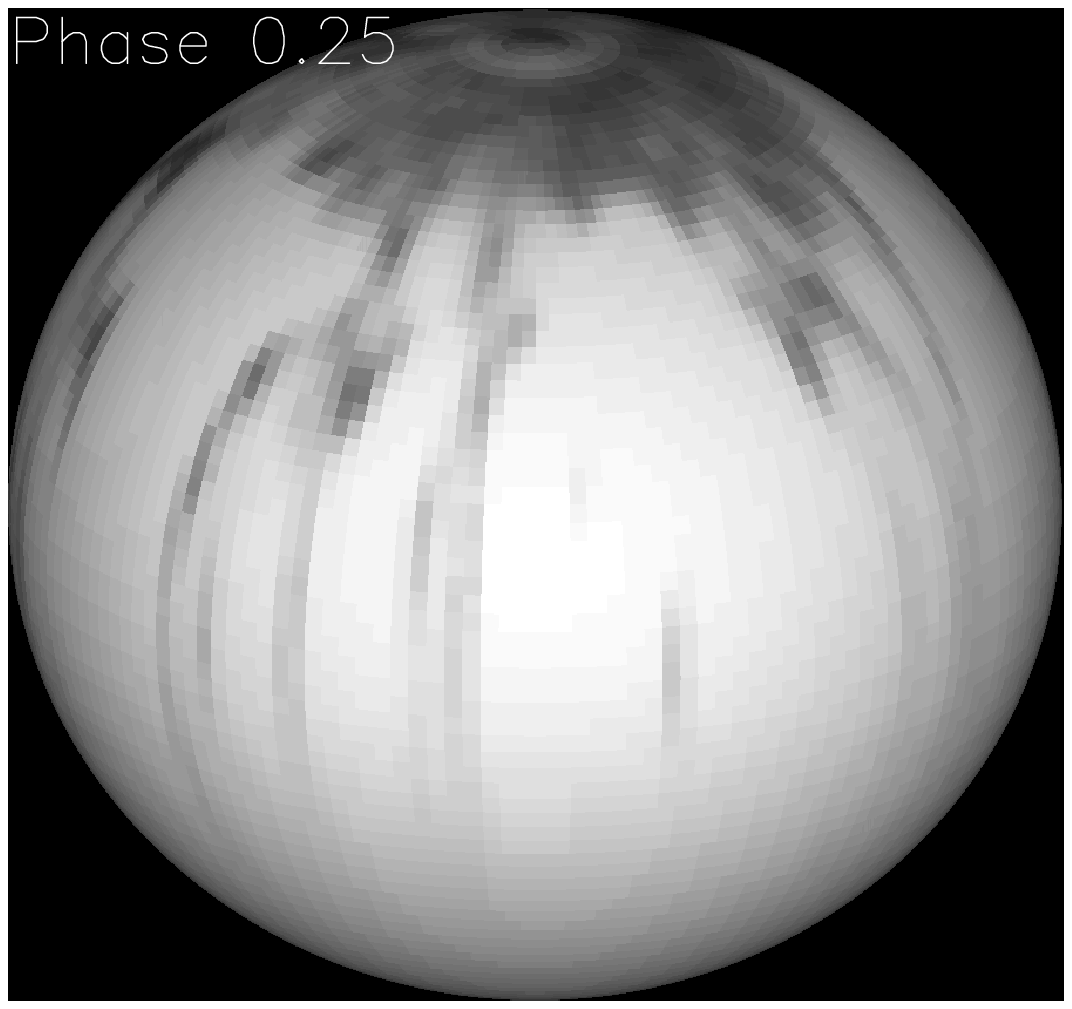} &
\includegraphics[height=40mm,angle=0,bbllx=100,bblly=267,bburx=390,bbury=559]{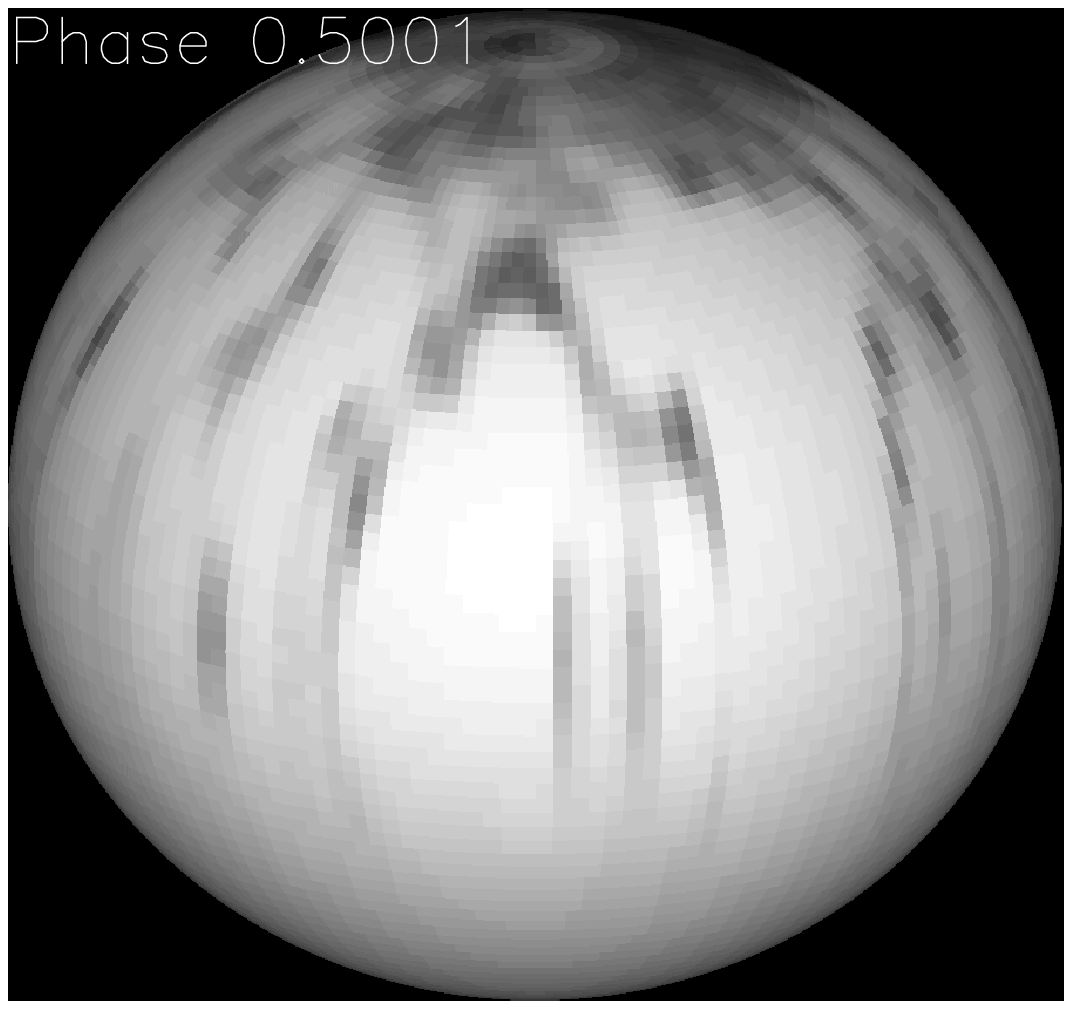} &
\includegraphics[height=40mm,angle=0,bbllx=100,bblly=267,bburx=390,bbury=559]{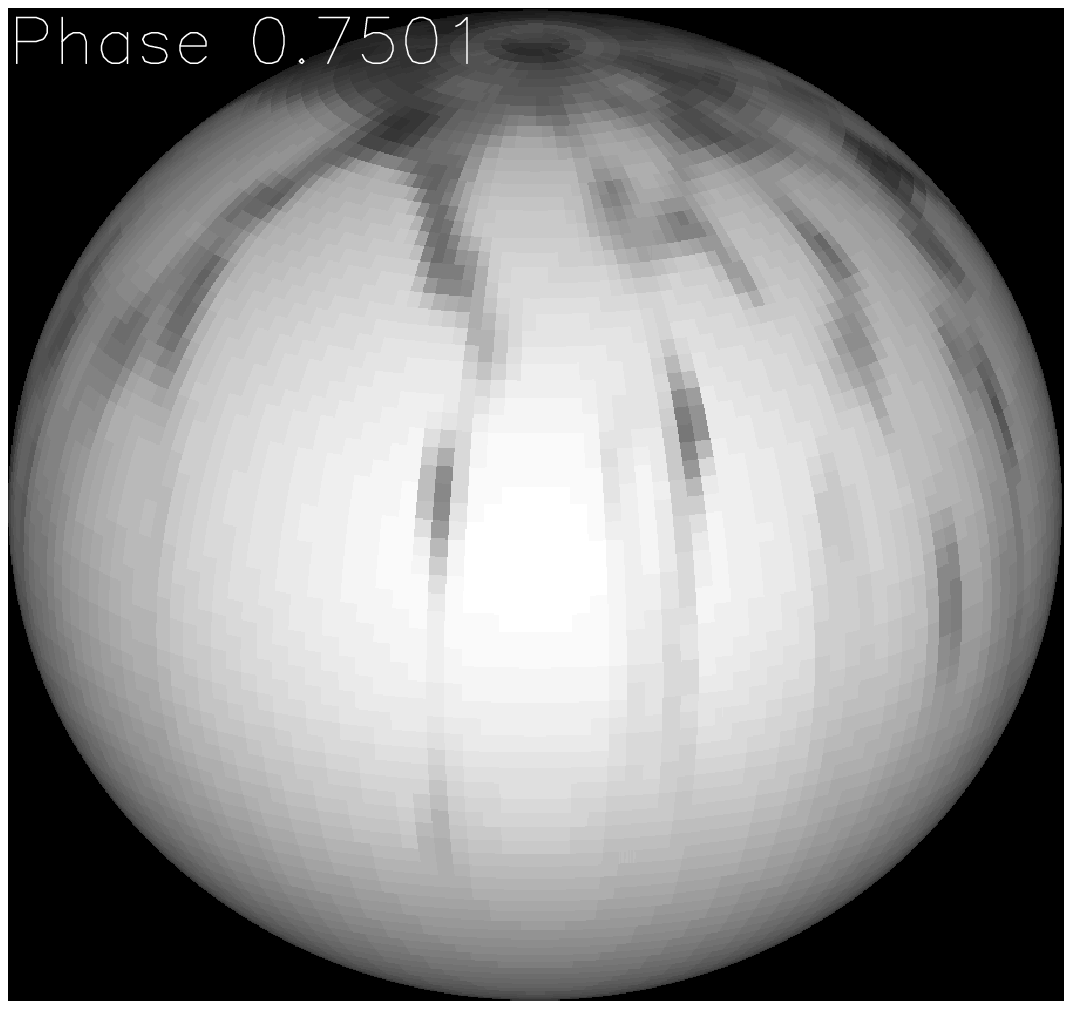} \\
\end{tabular}
\caption[Speedy reconstructed image]{Image reconstruction using 2002 July 18, 19 \& 23 combined. Differential rotation included. Top: Mercator projection, Bottom: Projected disc at 0.25 phase intervals.}
\protect\label{fig:speedy_alldata}
\end{figure*}

\section{THE IMAGES}
\protect\label{sec:images}

The reconstructed images of Speedy Mic are presented in Fig. \ref{fig:speedy_images} for each individual night of observations. All images are phased on the epoch HJD = 2452473.5, midnight on 2005 July 18. As expected from the numerous trails in Fig. \ref{fig:timeseries}, many starspot features are reconstructed at most latitudes and are found uniformly spread at all phases. The images from the nights of July 18 and 19 show many similarities, where the same spots are seen 2-3 rotations apart. By July 23 most spots at low and intermediate latitudes are also still visible whereas there may have been some evolution at higher latitudes.

\subsection{Image reliability}
\protect\label{sec:imagerel}

We have attempted to obtain some measure of the reliability of the reconstructed images by applying a Monte-Carlo bootstrap procedure \citep{efron79bootstrap, press92} to each night of observations. This involves generating a large number of data sets from our original data set and reconstructing images as follows. Firstly, we preserve the order of the rotation profiles. For the $n$ profiles, in a data set, we pick at random and with replacement $n$ profiles. If a given profile is chosen $m$ times, we divide the errors on that profile in the new data set by \mbox{$\sqrt{m}$}, whereas profiles not chosen are given very large error bars in order to exclude them from the timeseries. In this manner, we generated 100 data sets for each night of observations and reconstructed an image. This technique therefore requires no manipulation of the original profiles, and involves re-distributing the error weights only. As such, the amount of noise in the data is conserved. We can thus fit each data set to the same level of \chisq\ in order to obtain a reliable estimate of the uncertainties in reconstructed pixel values in the maps. 

An image with values representing the standard deviation for each pixel was generated and is shown in Fig. \ref{fig:speedy_stddev}. It is clear from these maps that most of the strong features are reliably re-constructed as the cores of the features show uncertainty in filling factor of typically 0.1. As expected, the elongated features at low latitude which typically show a spot occupancy of 0.1 also show high uncertainties in spot filling. These features are a reflection of the uncertainties in the rate at which starspots pass through the profile. The gradient is proportional to the sine of the latitude, $\theta$, and \mbox{$sin(\theta)$} is a rapidly changing function near the equator.

The main point of this exercise is to distinguish between evolution of spots and uncertainties in the reconstruction process. Focusing on the four spots seen at phase 0.8 to 0.9 and centred around latitude 60\degs in the July 18, we see that the arc-like structure joining the spots possess a greater uncertainty than the spots themselves. By July 19, there does appear to have been genuine evolution of these features, especially the two spots at phase at phase 0.8, although the reliability of the lower latitude spot of these two features is quite large on July 19. It is nevertheless clear that there are many features common to all images which makes us confident that we can reliably measure the differential rotation.

\subsection{Differential rotation}
\protect\label{sec:diffrot}

The large number of repeated features and high effective surface resolution (full width at half maximum of mapping profile is \hbox{7.69\ \kms}, corresponding to a resolution of 5.2\degs\ on the equator) make Speedy Mic an ideal target to attempt a measurement of differential rotation. By optimising the degree of fit of a solar-like differential rotation law (where \mbox{$\Omega(\theta) = \Omega_0 - \Delta\Omega\rm{sin}^2(\theta)$}, $\theta$ is the stellar latitude and $\Omega_0$ the equatorial rotation velocity) to the full data set, incorporating the full time series of data, we find that \hbox{\mbox{$\Omega_0 = 16.5361$}} \hbox{\mbox{$\pm 0.0006 $} \radday}\ and \hbox{\mbox{$\Delta\Omega = 0.03295 \pm 0.00294$ }}. This equates to an equator-pole-lap time of 191 $\pm$ 17 days. The result reported here differs slightly from the value of $199 \pm 13$ days given in \citet{barnes05diffrot}, but agrees within the uncertainties. The only difference here is that we have re-determined the optimal inclination angle, and use 70\degs\ rather than 55\degs.

\section{Discussion \& Conclusion}
\protect\label{sec:discussion}

The images of Speedy Mic reveal it to be one of the most active stars to have been imaged to date. This is largely due to the high S/N ratio of the data combined with high spectral resolution of $\sim$5\degs\ at the equator.  Unfortunately S/N variations of the data and a lower degree of phase coverage on July 23 make comparison of some features difficult or impossible. In Fig. \ref{fig:speedy_alldata} we have reconstructed an image of Speedy Mic using all nights of data and the optimised differential rotation shear.

\subsection{Spot distribution and evolution}
\protect\label{sec:evolution}

In order to obtain a better impression of the surface distribution of spots, we have also plotted the projected disc of Speedy Mic at four key phases. This gives a clearer representation of the polar regions which are severely distorted in the Mercator projections. The plots make it clear that there is in fact no single uniform polar spot, as is often seen on late-type rapidly rotating stars (e.g. see \citealt{strassmeier02survey}). Except for the large de-centred spot/spot-group at phase \hbox{0.0\, -\, 0.25}, the high latitudes and the polar regions appear to comprise a number of resolved closely packed spots. A similar de-centred spot was found on the young $\alpha$ Persei cluster G dwarf, He 520 \citep{barnes98aper, barnes01aper}. Whether a stellar magnetic cycle is responsible for the change in the morphology of spots in the polar region remains highly speculative. If meridional flows however are responsible for carrying erupted magnetic flux to high latitudes \citep{mackay04}, a persistent active region at a given longitude may be responsible for the appearance of a decentred spot.

The images are remarkably similar to the images presented in B01. Observations were much more sparse during the 1998 observations, and were made using more than one telescope and instrumental setup. Nevertheless, the uniform distribution of spots at all latitudes is found in both sets of observations. The increased filling at high latitudes in the 2002 data may be partly real, but may also be the consequence of using a slightly higher inclination for the image reconstructions (see B01).

Despite its rapid rotation, examination of the images in Fig. \ref{fig:speedy_images} shows that many features are repeated on both {one} and {five} day timescales. The spots may erupt in groups which are centred at phases of approximately 0.1, 0.3 and 0.5. Between phases 0.5 and 1.0, such grouping is less clear. We caution that one should not be tempted to think that the smeared arcs (see \S \ref{sec:images}) in Fig. \ref{fig:speedy_stddev} connect spots together in a group. These arcs are regions of greatest uncertainty, and probably a consequence of the maximum entropy reconstruction process.

Since observations of \speedy\ were made by WSW05 between two and three weeks after those presented here, we are in a position to examine evolution of surface features on this timescale. The observations presented by WSW05 were made with the UV-Visual \'{E}chelle Spectrograph at the 8-m Very Large Telescope, Kueyen. Deconvolution over a single spectral order enabled single profiles with S/N of several hundred to be obtained, and the images were derived using the updated {\sc CLEAN}-like algorithm, first applied to observations of the K0 dwarf AB Dor \citep{kurster93,kurster94abdor}. { The deconvolution technique applied by WSW05 differs from that used in this publication in that a template spectrum is iteratively convolved with an adjustable rotation profile, or broadening function, until an acceptable fit to the observed spectrum is obtained. This potentially enables a better fit to the observed spectrum to be derived as compared with the least squares deconvolution technique. The template spectrum method may be more efficient in boosting the effective S/N ratio than the least squares method, but the deformed rotation profile will be effectively the same in both instances. WSW05 applied this method to two independent single \'{e}chelle orders, yielding profiles with S/N ratios of 300\,-\,500. 

Whereas maximum-entropy regularisation is a continuous optimisation problem, giving the smoothest possible image (i.e. the one containing the least amount of information), the CLEAN-like approach attempts a discrete approximation of the line profile deformation (i.e. the starspots). Hence discrete filling factors are reconstructed, and it is argued that penumbral regions are more accurately represented using this method. \cite{kurster93} showed that CLEAN-like and maximum entropy reconstructions produce similar maps, but that the appearance is more `ragged' using the former method. This certainly appears to be the case when we compare our images with those of WSW05. However, it may be difficult to compare the two sets of images on such a small spatial scale since the S/N ratio of our data after deconvolution is nearly an order of magnitude greater as a result of using a larger spectral range. At S/N ratios of several thousand, the data should constrain the images very well with the image entropy playing a smaller role in the reconstruction. As already discussed, the smoothing in our images must be largely due to the maximum entropy regularisation. It thus seems probable that we are not making the most of the high quality deconvolved profiles. A re-examination of regularisation techniques where high S/N ratio data are involved may be appropriate, but is beyond the scope of this paper.}

We find a remarkable correlation between the spot groups found in the present images and the image derived for 2002 August 02 by WSW05 (see Fig. 6, 6400\AA\ image). We suggest that WSW05 find the same spot groups at longitudes of approximately 80\degs, 150\degs, 220\degs, \& 280\degs. Since our data are phased on a different epoch (see \S \ref{sec:images}) to those of WSW05, we find that our phase 0.0 corresponds to phase 0.24 in the WSW05 images. The longitudes of the spot groups in WSW05 correspond to phases 0.55, 0.34, 0.15 and 0.98 on our maps, and are remarkably close to the locations described above, of 0.5, 0.3 and 0.1. The group at 0.98 may correspond to the group found at phase $\sim 0.9$ in this work.

By 2005 August 7 (WSW05), the spot groups have evolved beyond recognition, apart from the phase 0.34 group which has become more extensive. The images presented here and in WSW05 allow us to place a lower limit on the lifetimes of individual spots at five days. By contrast, most of the larger spot groups persist for at least two weeks on Speedy Mic, with only one group being present over a three week period. Unfortunately we cannot make any estimate of the real duration of spot groups without a longer span of observations.

The two week timebase of the the data sets enables us to make an approximate verification of the differential rotation derived in this paper. The shift in the longitude position of spot groups is $\sim$0.05 (18\degs), so the same features are appearing slightly later in phase by August 02. If we make the approximation that the spot groups are centred at latitude 45\degs, then the differential rotation derived in \S \ref{sec:diffrot} implies $\Delta\Omega$ = 0.016, giving a phase shift of \hbox{0.0026 day$^{-1}$}. Over the 15 day span from July 18 - August 2, we therefore expect a total phase shift of 0.039, which is in agreement with our estimate of 0.05.

\subsection{Comparison with other objects}
\protect\label{sec:comparison}

Other stars with similar spectral type such as AB Dor and LQ Hya (K0V; \citealt{donati03imaging}), PZ Tel (K1V; \citealt{barnes00pztel}) and LO Peg (K5V; \citealt{lister99lopeg, barnes05lopeg}), reveal very similar distributions of spots. Generally these stars exhibit either several large spots (e.g. LQ Hya and LO Peg) or several spot groups distributed relatively uniformly in longitude and latitude. In addition to S/N ratio variations, it is clear that we must take into account the projected rotation velocities for each object when considering the degree of surface detail recovered. For \hbox{LQ Hya}, \vsini\ = 26 \kms\ and for \hbox{LO Peg}, \vsini\ = 65 \kms (at approximately 2/3 the instrumental resolution of the other objects). Both these objects exhibit large spots which no doubt comprise smaller unresolved features. Progressively more detail is resolved as we increase the projected rotation velocity, from PZ Tel \hbox{(\vsini = 68\kms)}, through AB Dor (\vsini\ = 90\kms), to Speedy Mic (\vsini\ = 132\kms). 

This resolution issue will also affect how much structure is resolved in the poles and whether the polar spot appears solid, since the spatial resolution at high latitudes is lower. It is difficult to see a trend however since the two most extreme cases in the sample (in terms of rotation rate) possess the smallest (LQ Hya) or least extensive (Speedy Mic) polar caps. Given the long term nature of observations for AB Dor, and the absence of polar spot in 1988 \citep{kurster94abdor}, it seems most likely that magnetic cyclic activity affects the size and presence of a polar spot. It is difficult however to determine whether \speedy\ exhibits cyclic activity, and whether the high latitude structure has changed significantly since the 1998 observations. Reconstructing the image with an axial inclination of 70\degs\ results in an increase in the strength of the high latitude features. An effective (fractional) spot area of $\sim0.05$ at low-intermediate latitudes and $\sim0.15$ at high latitudes are found (Fig. \ref{fig:speedy_images}, right panels). If the present images are reconstructed at 55\degs\ it is found that the effective spot area as a function of latitude varies little, peaking at around 0.075. The images then closely resemble those found in B01.

\subsection{Prospects}
\protect\label{sec:prospects}

It is difficult to state whether or not a shift in latitude of the starspot groups has occurred over the timebase of observations. Simulations which focus on the role of enhanced magnetic flux emergence on the subsequent transport of flux at the surface indicate that the mixed polarities seen on the K0 dwarf, AB Dor \citep{donati03imaging}, can be explained by meridional flows \citep{schrijver01polar,mackay04}. If the 100 \ms\ flows implied by these simulations are to be tested, daily observations over several days must be carried out. By investigating the possible change in the rate at which spots pass through the extent of the rotation profile in data space, it may be possible to obtain an estimate of the meridional flow patterns. Enhanced meridional flows may also be responsible for the presence of the polar cap.

Unfortunately Speedy Mic is too faint for spectropolarimetric observations which may allow us to obtain some indication of the nature of the polarities at high latitudes. Since the contrast is so high in the optical between spot and photosphere, we would not expect to obtain any information from the spots themselves. We must turn to the next generation of infrared spectrographs which will allow us to use lines which only form in the spots to measure the polarities of the spots themselves. In the near infrared at $\sim1.7\mu m$, the blackbody flux ratio is of order 2, whereas it is closer to 12 in the optical.

\section{ACKNOWLEDGMENTS}

The data in this paper were reduced using {\sc starlink} software packages. This research made use of Simbad/Aladin (http://cdsweb.u-strasbg.fr). Thanks to { Ron Hilditch,} Tim Lister and Nick Dunstone for helpful discussions throughout the course of this work.

\bibliographystyle{mn2e}
\bibliography{iau_journals,master,ownrefs}

\end{document}